\documentclass{article}
\newcommand {\bl}{\begin{list}{}{\leftmargin 2em}}
\newcommand {\ite}{\item{}\hspace{-2em}}
\newcommand {\el}{\end{list}}
\newcommand {\beq}{\begin{equation}}
\newcommand {\eeq}{\end{equation}}

\newcommand{\bm}[1]{\mbox{\boldmath $#1$}}

\newcounter{saveeqn}
\newcommand{\alpheqn}{\setcounter{saveeqn}{\value{equation}}%
\stepcounter{saveeqn}\setcounter{equation}{0}%
\renewcommand{\theequation}{\mbox{\arabic{saveeqn}\alph{equation}}}}
\newcommand{\reseteqn}{\setcounter{equation}{\value{saveeqn}}%
\renewcommand{\theequation}{\arabic{equation}}}

\newcounter{EQ1}

\usepackage{lineno}

\usepackage[dvips]{graphicx, color}
\usepackage{exscale, latexsym, makeidx, newlfont}
\usepackage{amsfonts, amsbsy}
\usepackage[mathscr]{euscript}
\begin{document}

\begin{center}
\section*{\bf Akaike's Bayesian information criterion (ABIC) or not ABIC for geophysical inversion}
\end{center}

\vspace{5mm}

\hspace*{-7mm} Peiliang Xu
\\ Disaster Prevention Research Institute, Kyoto University, Uji,  Kyoto 611-0011, Japan
\\ email: pxu@rcep.dpri.kyoto-u.ac.jp
\vspace{5mm}

\begin{center}
 \parbox{15cm}{
{\bf Abstract:} {\sl Akaike's Bayesian information criterion (ABIC) has been widely used
in geophysical inversion and beyond. However, little has been done to investigate its
statistical aspects. We present an alternative derivation of the marginal distribution of
measurements, whose maximization directly leads to the invention of ABIC by Akaike. We
show that ABIC is to statistically estimate the variance of measurements and the prior
variance by maximizing the marginal distribution of measurements. The determination of
the regularization parameter on the basis of ABIC is actually equivalent to estimating
the relative weighting factor between the variance of measurements and the prior variance
for geophysical inverse problems. We show that if the noise level of measurements is
unknown, ABIC tends to produce a substantially biased estimate of the variance of
measurements. In particular, since the prior mean is generally unknown but arbitrarily
treated as zero in geophysical inversion, ABIC does not produce a reasonable estimate for
the prior variance either.
\vspace{1mm} \\
{\bf Keywords:} Akaike's Bayesian information criterion, marginal likelihood functions,
prior variance, regularization parameter, variance of measurements. } }
\end{center}

\parindent 4mm

\section{Introduction}
A linear inverse ill-posed problem can often be written symbolically as the following
linear model:  \beq \label{IPModel} \mathbf{y} = \mathbf{A} \bm{\beta} + \bm{\epsilon},
\eeq where $\mathbf{y}$ is an $(n\times 1)$ vector of measurements, $\mathbf{A}$ is the
deterministic coefficient matrix, which is assumed to be theoretically of full column
rank but with singular values close to zero, $\bm{\beta}$ is a $(t\times 1)$ vector of
unknown parameters to be estimated, and the random error vector $\bm{\epsilon}$ is of
zero mean and variance-covariance matrix $\mathbf{W}\sigma^2$, $\mathbf{W}$ is an
$(n\times n)$ (positive definite) weighting matrix. If the least squares (LS) method is
applied to (\ref{IPModel}), we have the weighted LS solution of $\bm{\beta}$ below: \beq
\label{LSEst} \hat{\bm{\beta}} = ( \mathbf{A}^T\mathbf{WA})^{-1} \mathbf{A}^T\mathbf{W}
\mathbf{y}. \eeq Although $\hat{\bm{\beta}}$ is unbiased, it can become highly unstable
and practically meaningless physically, because the normal matrix
$\mathbf{A}^T\mathbf{WA}$, often denoted by $\mathbf{N}=\mathbf{A}^T\mathbf{WA}$, is of
almost zero eigenvalues (see e.g., Phillips 1962; Tikhonov 1963; Tikhonov and Arsenin
1977).

To obtain a mathematically and/or physically meaningful solution to inverse ill-posed
problems (\ref{IPModel}), we can either limit ourselves to a sub-space spanned by the
vectors corresponding to sufficiently large eigenvalues (see e.g., Xu 1998) or add a
positive (semi-)definite matrix, say $\mathbf{W}_{\beta}\kappa$, to the normal matrix
$\mathbf{N}$ such that the addition of both matrices avoids sufficiently small
eigenvalues, where $\mathbf{W}\!_{\beta}$ is positive (semi-)definite and $\kappa$ is a
positive scalar. This latter method is well known either as ridge regression in the
statistical literature (see e.g., Hoerl and Kennard 1970; Vinod and Ullah 1981; Xu 1992a;
Xu and Rummel 1994) or regularization in the literature of mathematics and applied
sciences (see e.g., Tikhonov 1963; Tikhonov and Arsenin 1977), with the solution being
given as follows: \beq \label{RegularizedSolut} \hat{\bm{\beta}}_r = (
\mathbf{A}^T\mathbf{WA}+\kappa \mathbf{W}\!_{\beta})^{-1} \mathbf{A}^T\mathbf{W}
\mathbf{y}, \eeq which will be called a regularized solution in this letter, for
simplicity but without loss of confusion.

Determination of the regularization parameter $\kappa$ is of crucial importance in
regularizing the inverse ill-posed model (\ref{IPModel}). If $\kappa$ is too small, the
regularized solution $\hat{\bm{\beta}}_r$ of (\ref{RegularizedSolut}) may still remain
unstable; however, if it is chosen too large, $\hat{\bm{\beta}}_r$ will become
over-smoothed. In particular, if $\kappa$ tends positively to infinity,
$\hat{\bm{\beta}}_r$ will shrink towards zero and the measurements $\mathbf{y}$ play no
role at all in retrieving the information on $\bm{\beta}$. There are a number of methods
to determine the regularization parameter $\kappa$, depending on how the regularized
solution $\hat{\bm{\beta}}_r$ of (\ref{RegularizedSolut}) is interpreted. The view points
can be either presented in terms of frequentist, under the Bayesian framework, or even
intuitively in terms of some norms of residuals of measurements and/or estimates of
$\bm{\beta}$.

From the frequentist point of view, $\hat{\bm{\beta}}_r$ of (\ref{RegularizedSolut}) has
been known as a biased estimator of $\bm{\beta}$. As a result, the regularization
parameter has been mainly motivated to compromise between noise amplification and bias
such as the criterion of mean squared errors (see e.g., Hoerl and Kennard 1970; Vinod and
Ullah 1981; Xu 1992a; Xu and Rummel 1994) or predict measurements such as generalized
cross-validation and maximum likelihood (see e.g., Golub et al. 1979; Wahba 1985; Xu
2009). Norms of residuals and parameters have played an important role in choosing the
regularization parameter. Very often, one can either determine the regularization
parameter, given a fixed level of noise for the residual norm (see e.g., Miller 1970;
Tikhonov and Arsenin 1977; Morozov 1984) or by finding a balance between some norms of
residuals and parameters (see e.g., Hansen and O'Leary DP 1993).

The inverse ill-posed model (\ref{IPModel}) has also been solved under the Bayesian
framework (see e.g., Akaike 1980; Tarantola 1987). Assuming that there exists prior
information on the parameters $\bm{\beta}$ in terms of the first and second (central)
moments, namely, \beq \label{PriorB} E(\bm{\beta}) = \bm{\mu}, \hspace{2mm} D(\bm{\beta})
= \mathbf{W}\!_{\beta}^{-1}\sigma_{\beta}^2, \eeq where the prior vector $\bm{\mu}$ is
given, $D(\cdot)$ stands for variance operator, $\mathbf{W}\!_{\beta}$ is a positive
definite matrix and the scalar $\sigma_{\beta}^2$ is given as well. Then applying
stochastic inference to inverse problems (\ref{IPModel}) with prior information
(\ref{PriorB}) results in the following estimator: \beq \label{StochasticI}
\hat{\bm{\beta}}_i = \left( \frac{\mathbf{A}^T\mathbf{WA}}{\sigma^2}+
\frac{\mathbf{W}\!_{\beta}}{\sigma_{\beta}^2}\right)^{-1} \left(
\frac{\mathbf{A}^T\mathbf{W} \mathbf{y}}{\sigma^2} +
\frac{\mathbf{W}\!_{\beta}\bm{\mu}}{\sigma_{\beta}^2} \right). \eeq If the prior values
$\bm{\mu}$ are equal to zero, namely, $\bm{\mu}=\mathbf{0}$, and denoting
$\kappa=\sigma^2/\sigma_{\beta}^2$, then the stochastic inference $\hat{\bm{\beta}}_i$ of
(\ref{StochasticI}) {\em formally} becomes the regularized or biased solution
$\hat{\bm{\beta}}_r$ of (\ref{RegularizedSolut}).

An alternative use of prior information to solve inverse problems (\ref{IPModel}) is
based on full Bayesian inference, which requires the assumption of distributions of both
measurements and prior data. As in the case of regularization and/or stochastic
inference, proper choice of the regularization parameter is crucial in Bayesian
inversion. Akaike (1980) proposed a Bayesian information criterion to determine the
regularization parameter, which has since been widely applied in geophysical inversion.
However, little has been known about statistical aspects of Akaike's Bayesian information
criterion (ABIC) for ill-posed inverse problems. The purpose of this paper is primarily
to investigate statistical performances of ABIC. We will provide an alternative
representation of the marginal distribution of measurements. We will show that the ABIC
determination of regularization parameter is statistically equivalent to estimating the
variance $\sigma^2$ of measurements and the prior variance $\sigma_{\beta}^2$. We will
also prove in this section that if the prior mean $\bm{\mu}$ is unknown but arbitrarily
treated as zero, as almost always in the case of practical geophysical inversion, the
estimate of $\sigma^2$ will be significantly biased, which will further affect the
estimate of $\sigma_{\beta}^2$ and thus further the determination of the regularization
parameter. The theoretical results will then be summarized in the concluding remarks.

\section{ABIC and its statistical aspects}
The measurement and prior distributions are usually assumed to be normal and respectively
given as follows: \alpheqn \beq \label{DATAPdf} f_y(\mathbf{y}/\bm{\beta}, \sigma^2) =
\frac{\sqrt{\textrm{det}(\mathbf{W})} }{(2\pi)^{n/2}\sigma^n} \exp\left\{
-\frac{1}{2\sigma^2}( \mathbf{y}-\mathbf{A}\bm{\beta})^T\mathbf{W}(
\mathbf{y}-\mathbf{A}\bm{\beta})\right\}, \eeq  and \beq \label{PriorPdf}
\pi\!_{\beta}(\bm{\beta}/ \sigma_{\beta}^2) =
\frac{\sqrt{\textrm{det}(\mathbf{W}\!_{\beta})} }{(2\pi)^{t/2}\sigma_{\beta}^t}
\exp\left\{ -\frac{1}{2\sigma_{\beta}^2}( \bm{\beta} - \bm{\mu} )^T\mathbf{W}\!_{\beta}(
\bm{\beta} - \bm{\mu})\right\}, \eeq\reseteqn\setcounter{EQ1}{\value{equation}}as can be
found, for example, in Zellner (1971) and Akaike (1980), where $\textrm{det}(\cdot)$
stands for the determinant of a square matrix. Then the joint distribution density of
both $\mathbf{y}$ and $\bm{\beta}$ is given by \beq \label{jointYBeta}
f(\mathbf{y},\bm{\beta}/\sigma^2,\sigma_{\beta}^2) = f_y(\mathbf{y}/\bm{\beta},
\sigma^2)\pi\!_{\beta}(\bm{\beta}/ \sigma_{\beta}^2). \eeq

Bayesian inference on $\bm{\beta}$ has been done on the basis of the following posterior
distribution of $\bm{\beta}$ given the measurements $\mathbf{y}$: \begin{eqnarray}
\label{postPDF} \pi(\bm{\beta}/\mathbf{y},\sigma^2,\sigma_{\beta}^2) & = &
\frac{f(\mathbf{y},\bm{\beta}/\sigma^2,
\sigma_{\beta}^2)}{m(\mathbf{y}/\sigma^2,\sigma_{\beta}^2)} \nonumber \\
 & = & \frac{f_y(\mathbf{y}/\bm{\beta},
\sigma^2)\pi\!_{\beta}(\bm{\beta}/
\sigma_{\beta}^2)}{m(\mathbf{y}/\sigma^2,\sigma_{\beta}^2)}, \end{eqnarray} where
$\pi(\bm{\beta}/\mathbf{y},\sigma^2,\sigma_{\beta}^2)$ stands for the posterior
distribution of $\bm{\beta}$ given $\mathbf{y}$, and
$m(\mathbf{y}/\sigma^2,\sigma_{\beta}^2)$ is the marginal distribution of $\mathbf{y}$,
which is defined and given as follows: \beq \label{marginalY}
m(\mathbf{y}/\sigma^2,\sigma_{\beta}^2) = \int
f(\mathbf{y},\bm{\beta}/\sigma^2,\sigma_{\beta}^2)  d\bm{\beta}. \eeq As a result, one
can derive the Bayesian  (posterior) estimate, either by maximizing or computing the
posterior mean from the posterior distribution
$\pi(\bm{\beta}/\mathbf{y},\sigma^2,\sigma_{\beta}^2)$ of (\ref{postPDF}), which is
denoted by $\hat{\bm{\beta}}_b$ and simply written below \begin{eqnarray}
\label{BayesianI} \hat{\bm{\beta}}_b & = & \left(
\frac{\mathbf{A}^T\mathbf{WA}}{\sigma^2}+
\frac{\mathbf{W}\!_{\beta}}{\sigma_{\beta}^2}\right)^{-1} \left(
\frac{\mathbf{A}^T\mathbf{W} \mathbf{y}}{\sigma^2} +
\frac{\mathbf{W}\!_{\beta}\bm{\mu}}{\sigma_{\beta}^2} \right),  \nonumber \\
 & = & (\mathbf{A}^T\mathbf{WA}+\kappa \mathbf{W})^{-1} (\mathbf{A}^T\mathbf{W}
 \mathbf{y} + \kappa \mathbf{W}\!_{\beta} \bm{\mu}), \end{eqnarray} (see e.g., Zellner
1971; Akaike 1980). It is obvious that both stochastic and Bayesian inferences result in
the same estimator under the assumption of normal distributions by comparing
$\hat{\bm{\beta}}_i$ of (\ref{StochasticI}) with $\hat{\bm{\beta}}_b$ of
(\ref{BayesianI}). The Bayesian posterior estimator (\ref{BayesianI}) can also be
equivalently rewritten as \beq \label{BayesianIWmean} \hat{\bm{\beta}}_b  =  \left(
\frac{\mathbf{A}^T\mathbf{WA}}{\sigma^2}+
\frac{\mathbf{W}\!_{\beta}}{\sigma_{\beta}^2}\right)^{-1} \left(
\frac{\mathbf{A}^T\mathbf{WA}\hat{\bm{\beta}} }{\sigma^2} +
\frac{\mathbf{W}\!_{\beta}\bm{\mu}}{\sigma_{\beta}^2} \right), \eeq indicating that
$\hat{\bm{\beta}}_b$ is actually the weighted mean of the LS estimate $\hat{\bm{\beta}}$
and the prior mean $\bm{\mu}$. If the prior distribution is too poor to be useful, or
more precisely, $\sigma_{\beta}^2\Longrightarrow \infty$, then $\hat{\bm{\beta}}_b$
becomes $\hat{\bm{\beta}}$. However, for an ill-posed inverse problem, the LS estimate
$\hat{\bm{\beta}}$ can be too poor in accuracy and the prior information will constrain
$\hat{\bm{\beta}}_b$ even for a large value of $\sigma_{\beta}^2$.

If $\bm{\mu}=\mathbf{0}$ in (\ref{BayesianI}), the corresponding Bayesian estimator
 is then denoted by $\hat{\bm{\beta}}_{b0}$ and (\ref{BayesianI}) becomes
\begin{eqnarray} \label{BayesianISimple} \hat{\bm{\beta}}_{b0} & = & \left(
\frac{\mathbf{A}^T\mathbf{WA}}{\sigma^2}+
\frac{\mathbf{W}\!_{\beta}}{\sigma_{\beta}^2}\right)^{-1} \frac{\mathbf{A}^T\mathbf{W}
\mathbf{y}}{\sigma^2} \nonumber \\
 & = & (\mathbf{A}^T\mathbf{WA}+\kappa \mathbf{W}\!_{\beta})^{-1} \mathbf{A}^T\mathbf{W}
\mathbf{y}, \end{eqnarray} which essentially turns out to be $\hat{\bm{\beta}}_r$ of
(\ref{RegularizedSolut}).

{\em Remark 1}: For a geophysical inverse problem, the geophysical signals we are seeking
from measurements $\mathbf{y}$ are not equal to zero. In other words, it is generally not
fair/reasonable to assume $\bm{\mu}=\mathbf{0}$; otherwise, we have no reasons to collect
the data $\mathbf{y}$, as explained in the case of satellite gravimetry (Xu and Rummel
1994). By setting $\bm{\mu}=\mathbf{0}$ for geophysical inverse problems, the Bayesian
estimator (\ref{BayesianISimple}) is biased, as correctly pointed out by Akaike (1980)
(see also Xu 1992b). From this point of view, it may become easily understandable why the
regularized solution $\hat{\bm{\beta}}_r$ of (\ref{RegularizedSolut}) has been called a
biased estimator from the frequentist point of view.

Realizing the importance of choosing the parameter $\kappa$, or equivalently $\sigma^2$
and $\sigma_{\beta}^2$, Akaike (1980) proposed a criterion by maximizing the marginal
distribution of $\mathbf{y}$, {\em i.e.} $m(\mathbf{y}/\sigma^2,\sigma_{\beta}^2)$ of
(\ref{marginalY}) to choose $\kappa$, which is known as {\em Akaike's Bayesian
information criterion} or simply ABIC. ABIC has since been widely applied to geophysical
inverse problems, as can be seen, for example, in Tamura et al. (1991), Yabuki and
Matsu'ura (1992) and Fukuda and Johnson (2008), just to name a few of geophysical
applications.

Although $m(\mathbf{y}/\sigma^2,\sigma_{\beta}^2)$ plays a key role in ABIC, it was
simply given in Akaike (1980) without providing any details of its derivation, which has
been further basically used verbatim in geophysical applications (see e.g., Tamura et al.
1991; Yabuki and Matsu'ura 1992). Given the normal distribution (\ref{DATAPdf}) of
measurements and the normal prior distribution (\ref{PriorPdf}), we have re-derived the
marginal distribution $m(\mathbf{y}/\sigma^2,\sigma_{\beta}^2)$ of the measurements
$\mathbf{y}$, which is simply given as follows: \begin{eqnarray} \label{marginalYFinal}
m(\mathbf{y}/\sigma^2,\sigma_{\beta}^2) & = & \int
f(\mathbf{y},\bm{\beta}/\sigma^2,\sigma_{\beta}^2)  d\bm{\beta} \nonumber \\
 & = & \frac{1}{(2\pi)^{n/2}\sqrt{\textrm{det}(\bm{\Sigma}_{py})}} \exp\left\{
-\frac{1}{2}( \mathbf{y}-\mathbf{A}\bm{\mu})^T\bm{\Sigma}_{py}^{-1}(
\mathbf{y}-\mathbf{A}\bm{\mu})\right\}, \end{eqnarray} where \beq \label{YTotalVC}
\bm{\Sigma}_{py} = \mathbf{W}^{-1}\sigma^2 +
\mathbf{A}\mathbf{W}\!_{\beta}^{-1}\mathbf{A}^T\sigma_{\beta}^2. \eeq The detailed
derivation of (\ref{marginalYFinal}) is given in appendix A. The univariate version of
(\ref{marginalYFinal}) can be found in Zellner (1971, p.28). The marginal distribution
$m(\mathbf{y}/\sigma^2,\sigma_{\beta}^2)$  appears to be formally different from the
corresponding distribution in Akaike (1980), they are essentially identical.

It has become clear that ABIC for geophysical inversion is the same as estimating the two
unknown variance components $\sigma^2$ and $\sigma_{\beta}^2$ or equivalently, the
unknown variance $\sigma^2$ and the relative weight $\kappa$, by maximizing the marginal
distribution (\ref{marginalYFinal}) of measurements. Mathematically, maximizing
$m(\mathbf{y}/\sigma^2,\sigma_{\beta}^2)$ of (\ref{marginalYFinal}) is equivalent to
minimizing
 \beq \label{minMarginalY} \textrm{min:} \hspace{2mm} \mathcal{L}(\sigma^2,\sigma_{\beta}^2)
 =  \ln\{\textrm{det}(\bm{\Sigma}_{py})\}
 + ( \mathbf{y}-\mathbf{A}\bm{\mu})^T\bm{\Sigma}_{py}^{-1}(
\mathbf{y}-\mathbf{A}\bm{\mu}). \eeq Since $\kappa=\sigma^2/\sigma_{\beta}^2$, the
objective function $\mathcal{L}(\sigma^2,\sigma_{\beta}^2)$ can also be rewritten in
terms of $\sigma^2$ and $\kappa$ as follows: \beq \label{marginalYFun}
\mathcal{L}(\sigma^2,\kappa) = n\ln(\sigma^2) + \ln\{\textrm{det}(\mathbf{E}_{py})\}
 + ( \mathbf{y}-\mathbf{A}\bm{\mu})^T\mathbf{E}_{py}^{-1}(
\mathbf{y}-\mathbf{A}\bm{\mu})/\sigma^2, \eeq where $$ \mathbf{E}_{py} = \mathbf{W}^{-1}
+ \mathbf{A}\mathbf{W}\!_{\beta}^{-1}\mathbf{A}^T/\kappa. $$ We should like to note that
$\mathcal{L}(\sigma^2,\kappa)$ of (\ref{marginalYFun}) is essentially the same as the
corresponding formula in Section 5 of Akaike (1980).

\subsection*{Case 1: both $\sigma^2$ and $\sigma_{\beta}^2$ unknown.}
Differentiating $\mathcal{L}(\sigma^2,\kappa)$ of (\ref{marginalYFun}) with respect to
$\sigma^2$ and letting it equal to zero, we have
$$ \frac{n}{\widehat{\sigma^2}} - \frac{( \mathbf{y}-\mathbf{A}\bm{\mu})^T\mathbf{E}_{py}^{-1}(
\mathbf{y}-\mathbf{A}\bm{\mu})}{(\widehat{\sigma^2})^2} = 0, $$ or equivalently, \beq
\label{sigmaEst} \widehat{\sigma^2} = (
\mathbf{y}-\mathbf{A}\bm{\mu})^T\mathbf{E}_{py}^{-1}( \mathbf{y}-\mathbf{A}\bm{\mu}) / n.
\eeq Given the prior variance $\sigma_{\beta}^2$ or the equivalent relative weight
$\kappa$, it is rather easy to prove that $\widehat{\sigma^2}$ of (\ref{sigmaEst}) is an
unbiased estimate of $\sigma^2$. Substituting (\ref{sigmaEst}) into (\ref{marginalYFun})
and neglecting a constant term, we have the ABIC for determining the relative weight (or
regularization parameter) $\kappa$ as follows: \beq \label{marginalYABIC}
\mathcal{L}(\kappa) = n\ln\{( \mathbf{y}-\mathbf{A}\bm{\mu})^T\mathbf{E}_{py}^{-1}(
\mathbf{y}-\mathbf{A}\bm{\mu})\} + \ln\{\textrm{det}(\mathbf{E}_{py})\}. \eeq The
likelihood function $\mathcal{L}(\kappa)$ consists of two parts: a positive definite
quadratic form of the predicted residuals $(\mathbf{y}-\mathbf{A}\bm{\mu})$ and their
cofactor matrix $\mathbf{E}_{py}$. It is easy to prove mathematically that the first part
of $\mathcal{L}(\kappa)$ increases with $\kappa$, while the second part decreases with
the increase of $\kappa$.

Although $\bm{\mu}$ is generally not equal to zero but unknown in geophysical inversion,
it is almost always replaced with a zero vector under the framework of Bayesian
geophysical inversion. In this case, by imposing $\bm{\mu}=\mathbf{0}$ (even if we know
it cannot not zero), both formulae (\ref{sigmaEst}) and (\ref{marginalYABIC}) become \beq
\label{sigmaEstYOnly} \widehat{\sigma^2_s} = \mathbf{y}^T\mathbf{E}_{py}^{-1} \mathbf{y}
/ n, \eeq and  \beq \label{marginalYOnlyABIC} \mathcal{L}_s(\kappa) = n\ln\{
\mathbf{y}^T\mathbf{E}_{py}^{-1} \mathbf{y}\} + \ln\{\textrm{det}(\mathbf{E}_{py})\},
\eeq respectively.

Given a value $\kappa$, applying the expectation operator to (\ref{sigmaEstYOnly}) yields
\beq \label{BiasedSigma} E(\widehat{\sigma^2_s}) = E(\mathbf{E}_{py}^{-1}
\mathbf{y}\mathbf{y}^T)/n = \overline{\mathbf{y}}^T\mathbf{E}_{py}^{-1}
\overline{\mathbf{y}}/n + \textrm{tr}\{\mathbf{E}_{py}^{-1}\mathbf{W}^{-1}\}\sigma^2/n,
\eeq which clearly indicates that $\widehat{\sigma^2_s}$ of (\ref{sigmaEstYOnly}) can be
significantly biased from its true value $\sigma^2$, depending on the true values of
measurements, where $\overline{\mathbf{y}}$ stands for the vector of true values of
measurements $\mathbf{y}$. On the other hand, since both $\mathcal{L}(\kappa)$ of
(\ref{marginalYABIC}) and $\mathcal{L}_s(\kappa)$ of (\ref{marginalYOnlyABIC}) are
different only in whether the predicted residuals $( \mathbf{y}-\mathbf{A}\bm{\mu})$ or
the measurements $\mathbf{y}$ are used to compute the positive definite quadratic form.
Since $\mathbf{y}$ are generally larger than the residuals $(
\mathbf{y}-\mathbf{A}\bm{\mu})$, the relative weighting factor or regularization
parameter $\kappa$ is expected to be estimated with a bias. As a result, we expect that
the prior variance $\sigma_{\beta}^2$ will also be computed with a bias from both
$\widehat{\sigma^2_s}$ and $\kappa$.

\subsection*{Case 2: $\sigma^2$ known but $\sigma_{\beta}^2$ unknown.}
If $\sigma^2$ is given/known but $\sigma_{\beta}^2$ unknown, use of ABIC to determine the
regularization parameter is mathematically equivalent to finding the best estimate of the
prior variance $\sigma_{\beta}^2$ such that it maximizes the marginal distribution
$m(\mathbf{y}/\sigma^2,\sigma_{\beta}^2)$ of measurements to most favor the output of the
measurements $\mathbf{y}$. It is also equivalent to finding the optimal ABIC estimate of
$\sigma_{\beta}^2$ (or $\kappa$) that minimizes the following likelihood function: \beq
\label{marginalYBayesFun} \mathcal{L}_b(\kappa) = (
\mathbf{y}-\mathbf{A}\bm{\mu})^T\mathbf{E}_{py}^{-1}(
\mathbf{y}-\mathbf{A}\bm{\mu})/\sigma^2
 + \ln\{\textrm{det}(\mathbf{E}_{py})\}. \eeq As in the case of $\mathcal{L}(\kappa)$ in
(\ref{marginalYABIC}), the first term of $\mathcal{L}_b(\kappa)$ increases with $\kappa$,
while its second term with the cofactor matrix $\mathbf{E}_{py}$ decreases with the
increase of $\kappa$.

Since the prior mean $\bm{\mu}$ is practically unknown in geophysical inversion, it is
almost always treated as zero and the likelihood function $\mathcal{L}_b(\kappa)$ of
(\ref{marginalYBayesFun}) becomes \beq \label{marginalYOnlyBayesFun}
\mathcal{L}_{bs}(\kappa) = \mathbf{y}^T\mathbf{E}_{py}^{-1} \mathbf{y}/\sigma^2
 + \ln\{\textrm{det}(\mathbf{E}_{py})\}. \eeq Since $\mathbf{y}$ is generally expected to
  be much larger in size than the predicted residual vector
 $(\mathbf{y}-\mathbf{A}\bm{\mu})$, the optimal $\kappa$ from minimizing
 (\ref{marginalYOnlyBayesFun}) may tend to be smaller than that from minimizing
 (\ref{marginalYBayesFun}). Thus, arbitrarily assigning the prior mean $\bm{\mu}$ to zero
 would affect the ABIC estimate of $\sigma_{\beta}^2$.

\section{Concluding remarks}
Akaike's Bayesian information criterion (ABIC) was proposed by Akaike (1980) and has
since been widely applied in geophysical inversion. The ABIC method is to determine the
regularization parameter such that the marginal distribution of measurements is
maximized. In other words, the ABIC-based regularization parameter is optimally chosen to
most favor the output of the collected measurements. However, little has been done to
investigate its statistical aspects for geophysical inversion. We have presented an
alternative representation of the marginal distribution of measurements, which is the
starting point of ABIC. We have shown that ABIC for geophysical inverse problems is
statistically equivalent to estimating the variance of measurements and the prior
variance by maximizing the marginal distribution of measurements or minimizing the
likelihood function of the variance of measurements and the prior variance. If the prior
distribution is correct, the regularization parameter actually reflects the relative
weighting factor between these two variances. We have proved that if the noise level of
measurements is unknown, ABIC tends to produce a substantially biased estimate of the
variance of measurements. In particular, since the prior mean is generally unknown but
arbitrarily treated as zero in geophysical inversion, ABIC does not produce a reasonable
estimate for the prior variance either. In case of a given variance of measurements, the
determination of the regularization parameter on the basis of ABIC is mathematically
equivalent to estimating the prior variance for geophysical inverse problems. We may also
like to note that although ABIC maximizes the marginal distribution of measurements under
the Bayesian framework, it does not directly target at constructing a solution of high
quality in terms of mean squared errors.

{\bf Acknowledgements:} 
This work is partially supported by the National Natural Science Foundation of China
(41874012 and 41674013). 

\section*{References}
\bl
 \ite Akaike H (1980). Likelihood and the Bayes procedure. in: {\em Bayesian Statistics},
 Bernardo JM, de Groot MH, Lindley DV, Smith AFM, eds., University Press, Valencia,
 Spain,  pp.1-13.
 \ite Anderson TW, Rubin H (1956). Statistical inference in factor analysis. In {\em
 Proceedings of the third Berkeley symposium on mathematical statistics and probability},
 Vol.5, pp. 111-150.
  \ite Fukuda J, Johnson KM (2008). A fully Bayesian inversion for spatial distribution of
fault slip with objective smoothing. {\em Bull. seism. Soc. Am.}, {\bf 98}, 1128-1146.
  \ite Golub GM, Heath M, Wahba G (1979). Generalized cross-validation
 as a method for choosing a good ridge parameter. {\em Technometrics}, {\bf
 21}, 215-223.
  \ite Hansen PC, O'Leary DP (1993). The use of the L-curve in the
regularization of discrete ill-posed problems. {\em SIAM J. Sci. Comput.}, {\bf 14},
1487-1503.
 \ite Heiskanen W, Moritz H (1967). {\em Physical Geodesy}. Freeman, San
Francisco.
 \ite Hoerl AE, Kennard RW (1970). Ridge regression: Biased estimation for
nonorthogonal problems. {\em Technometrics}, {\bf 12}, 55-67.
 \ite Miller K (1970). Least squares methods for ill-posed problems with a
prescribed bound. {\em SIAM J. math. Anal.}, {\bf 1}, 52-74.
 \ite Morozov VA (1984). {\em Methods for Solving Incorrectly Posed Problems},
 Springer, Berlin.
 \ite Phillips DL (1962). A technique for the numerical solution of certain
 integral equations of the first kind. {\em J. Assoc. Comput. Mach.}, {\bf 9},
84-97. 
 \ite Reed GB (1973). Application of kinematical geodesy for determining the
short wave length components of the gravity field by satellite gradiometry. {\em Reports
of the Department of Geodetic Science}, No.201, Ph.D. dissertation, The Ohio State
University.
 \ite Rummel R (1986). Satellite gradiometry. In: {\em Mathematical and Numerical
Techniques in Physical Geodesy}, edited by H S\"{u}nkel, Springer, Berlin, pp.317-363.
 \ite Tamura Y, Sato T, Ooe M, Ishiguro M (1991). A procedure for tidal analysis with a Bayesian
information criterion. {\em Geophys. J. Int.}, {\bf 104}, 507-516.
 \ite Tarantola A (1987). {\em Inverse Problem Theory: Methods for Data Fitting
and Model Parameter Estimation}. Elsevier, Amsterdam.
 \ite Tikhonov AN (1963). Regularization of incorrectly posed problems, {\em
Soviet Math.}, {\bf 4}, 1624-1627. 
 \ite Tikhonov AN, Arsenin VY (1977). {\em Solutions of Ill-posed Problem}.
 John Wiley \& Sons, New York.
\ite Vinod H, Ullah A (1981). {\em Recent Advances in Regression}. Marcel Dekker, New
York.
  \ite Wahba G (1985). A comparison of GCV and GML for choosing the smoothing
 parameter in the generalized spline smoothing problem. {\em Ann. Statist.},
 {\bf 13}, 1378-1402.
  \ite Xu PL (1992a). Determination of surface gravity anomalies using gradiometric
observables.   {\em Geophys. J. Int.}, {\bf 110}, 321-332.
 \ite Xu PL (1992b). The value of minimum norm estimation of geopotential fields.
{\em Geophys. J. Int.}, 111, 170-178.
 \ite Xu PL (1998) Truncated SVD methods for linear discrete ill-posed problems.
{\em Geophys. J. Int.}, {\bf 135}, 505-514.
  \ite Xu PL (2009) Iterative generalized cross-validation for fusing
 heteroscedastic data of inverse ill-posed problems. {\em Geophys. J. Int.}, {\bf 179}, 182-200,
 doi: 10.1111/j.1365-246X.2009.04280.x
 \ite Xu PL, Rummel R (1994). A generalized ridge regression method with
applications in determination of potential fields. {\em manuscr. geodaetica}, {\bf 20},
8-20.
 \ite Yabuki T, Matsu'ura M (1992). Geodetic data inversion using a Bayesian
 information criterion for spatial distribution of fault slip. {\em Grophys. J. Int.},
 {\bf 109}, 363-375.
 \ite Zellner A (1971). {\em An Introduction to Bayesian Inference in Econometrics}.
 Wiley, New York.
\el

\section*{Appendix A: the derivation of the marginal distribution
$m(\mathbf{y}/\sigma^2,\sigma_{\beta}^2)$ of measurements $\mathbf{y}$ in
(\ref{marginalY}) } Based on the normal distributions (\ref{DATAPdf}) and
(\ref{PriorPdf}), we can write the major exponent part of the joint distribution
(\ref{jointYBeta}), denoted by $F(\mathbf{y},\bm{\beta}/\sigma^2,\sigma_{\beta}^2)$,  as
follows:
\begin{eqnarray} \label{QuadYBeta} F(\mathbf{y},\bm{\beta}/\sigma^2,\sigma_{\beta}^2) & = &
\frac{1}{\sigma^2}( \mathbf{y}-\mathbf{A}\bm{\beta})^T\mathbf{W}(
\mathbf{y}-\mathbf{A}\bm{\beta}) + \frac{1}{\sigma_{\beta}^2}( \bm{\mu} -\bm{\beta}
)^T\mathbf{W}\!_{\beta}( \bm{\mu} - \bm{\beta}) \nonumber \\
 & = & (\bm{\beta} - \hat{\bm{\beta}}_b)^T \left( \frac{\mathbf{A}^T\mathbf{WA}}{\sigma^2}+
\frac{\mathbf{W}\!_{\beta}}{\sigma_{\beta}^2}\right)(\bm{\beta} - \hat{\bm{\beta}}_b)
  + F_y(\mathbf{y}/\sigma^2,\sigma_{\beta}^2),\end{eqnarray} where \beq \label{QuadY}
 F_y(\mathbf{y}/\sigma^2,\sigma_{\beta}^2)  =
 \frac{1}{\sigma^2}( \mathbf{y}-\mathbf{A}\hat{\bm{\beta}}_b)^T\mathbf{W}(
\mathbf{y}-\mathbf{A}\hat{\bm{\beta}}_b) + \frac{1}{\sigma_{\beta}^2}( \bm{\mu} -
\hat{\bm{\beta}}_b )^T\mathbf{W}\!_{\beta}( \bm{\mu} - \hat{\bm{\beta}}_b), \eeq and
 $\hat{\bm{\beta}}_b$ has been given in (\ref{BayesianI}).

We now rewrite $F_y(\mathbf{y}/\sigma^2,\sigma_{\beta}^2)$ of (\ref{QuadY}) as follows:
\begin{eqnarray} \label{QuadYT1}
 F_y(\mathbf{y}/\sigma^2,\sigma_{\beta}^2) & = &
\frac{1}{\sigma^2}[( \mathbf{y}-\mathbf{A}\bm{\mu}) +
\mathbf{A}(\bm{\mu}-\hat{\bm{\beta}}_b)]^T\mathbf{W}[( \mathbf{y}-\mathbf{A}\bm{\mu}) +
\mathbf{A}(\bm{\mu}-\hat{\bm{\beta}}_b)] \nonumber \\
 &  & + \frac{1}{\sigma_{\beta}^2}(\bm{\mu}-
\hat{\bm{\beta}}_b  )^T\mathbf{W}\!_{\beta}( \bm{\mu} - \hat{\bm{\beta}}_b) \nonumber \\
 & = & ( \mathbf{y}-\mathbf{A}\bm{\mu})^T\frac{\mathbf{W}}{\sigma^2}(
 \mathbf{y}-\mathbf{A}\bm{\mu})+ ( \mathbf{y}-\mathbf{A}\bm{\mu})^T\frac{\mathbf{W}}{\sigma^2}( \bm{\mu} -
 \hat{\bm{\beta}}_b) \nonumber \\
 &   & + ( \bm{\mu} - \hat{\bm{\beta}}_b)^T\frac{\mathbf{W}}{\sigma^2}(
 \mathbf{y}-\mathbf{A}\bm{\mu}) \nonumber \\
 &   & + ( \bm{\mu} - \hat{\bm{\beta}}_b)^T\frac{\mathbf{W}}{\sigma^2}( \bm{\mu} -
 \hat{\bm{\beta}}_b) + ( \bm{\mu} - \hat{\bm{\beta}}_b)^T\frac{\mathbf{W}\!_{
 \beta}}{\sigma_{\beta}^2}( \bm{\mu} - \hat{\bm{\beta}}_b) \nonumber \\
 & = & ( \mathbf{y}-\mathbf{A}\bm{\mu})^T\frac{\mathbf{W}}{\sigma^2}(
 \mathbf{y}-\mathbf{A}\bm{\mu}) + ( \bm{\mu} - \hat{\bm{\beta}}_b)^T\left\{
 \frac{\mathbf{W}}{\sigma^2} + \frac{\mathbf{W}\!_{
 \beta}}{\sigma_{\beta}^2}\right\} ( \bm{\mu} - \hat{\bm{\beta}}_b) \nonumber \\
 &   & + ( \mathbf{y}-\mathbf{A}\bm{\mu})^T\frac{\mathbf{W}}{\sigma^2}( \bm{\mu} -
 \hat{\bm{\beta}}_b) + ( \bm{\mu} - \hat{\bm{\beta}}_b)^T\frac{\mathbf{W}}{\sigma^2}(
 \mathbf{y}-\mathbf{A}\bm{\mu}).
\end{eqnarray}

On the other hand, since the Bayesian estimator $\hat{\bm{\beta}}_b$ is derived from the
following normal equations: $$ \left[ \begin{array}{c} \mathbf{A} \\ \mathbf{I}
\end{array} \right]^T \left[ \begin{array}{cc} \mathbf{W}/\sigma^2 & \mathbf{0} \\
\mathbf{0} & \mathbf{W}\!_{ \beta}/\sigma_{\beta}^2 \end{array} \right] \left[
\begin{array}{c} \mathbf{y} - \mathbf{A}\hat{\bm{\beta}}_b  \\ \bm{\mu} -
\hat{\bm{\beta}}_b \end{array} \right] = \mathbf{0}, $$ or
 $$ \mathbf{A}^T\frac{\mathbf{W}}{\sigma^2} (\mathbf{y} - \mathbf{A}\hat{\bm{\beta}}_b) +
 \frac{\mathbf{W}\!_{ \beta}}{\sigma_{\beta}^2} (\bm{\mu} -
\hat{\bm{\beta}}_b) = \mathbf{0}, $$ which is equivalent to
 \beq \label{BayesianNormal} \mathbf{A}^T\frac{\mathbf{W}}{\sigma^2} (\mathbf{y} -
 \mathbf{A}\bm{\mu}) + \mathbf{A}^T\frac{\mathbf{W}}{\sigma^2}\mathbf{A} (\bm{\mu} -
\hat{\bm{\beta}}_b) +
 \frac{\mathbf{W}\!_{ \beta}}{\sigma_{\beta}^2} (\bm{\mu} -
\hat{\bm{\beta}}_b) = \mathbf{0}. \eeq Thus, we have \beq \label{BayesianIA} (\bm{\mu} -
\hat{\bm{\beta}}_b) = - \left[ \frac{\mathbf{A}^T\mathbf{W}\mathbf{A}}{\sigma^2} +
\frac{\mathbf{W}\!_{ \beta}}{\sigma_{\beta}^2} \right]^{-1}
\mathbf{A}^T\frac{\mathbf{W}}{\sigma^2} (\mathbf{y} - \mathbf{A}\bm{\mu}).
 \eeq

By substituting $(\bm{\mu} - \hat{\bm{\beta}}_b)$ of (\ref{BayesianIA}) into
(\ref{QuadYT1}) and after some slight rearrangement, we finally obtain
\begin{eqnarray} \label{QuadYFinal}
 F_y(\mathbf{y}/\sigma^2,\sigma_{\beta}^2) & = &
( \mathbf{y}-\mathbf{A}\bm{\mu})^T\frac{\mathbf{W}}{\sigma^2}(
 \mathbf{y}-\mathbf{A}\bm{\mu}) \nonumber \\
 &  & -( \mathbf{y}-\mathbf{A}\bm{\mu})^T \frac{\mathbf{W}}{\sigma^2} \mathbf{A}
 \left[ \frac{\mathbf{A}^T\mathbf{W}\mathbf{A}}{\sigma^2} +
\frac{\mathbf{W}\!_{ \beta}}{\sigma_{\beta}^2}
\right]^{-1}\mathbf{A}^T\frac{\mathbf{W}}{\sigma^2}( \mathbf{y}-\mathbf{A}\bm{\mu})
\nonumber \\
 & = & ( \mathbf{y}-\mathbf{A}\bm{\mu})^T\left\{ \frac{\mathbf{W}}{\sigma^2} -
  \frac{\mathbf{W}}{\sigma^2} \mathbf{A}
 \left[ \frac{\mathbf{A}^T\mathbf{W}\mathbf{A}}{\sigma^2} +
\frac{\mathbf{W}\!_{ \beta}}{\sigma_{\beta}^2}
\right]^{-1}\mathbf{A}^T\frac{\mathbf{W}}{\sigma^2} \right\} (
\mathbf{y}-\mathbf{A}\bm{\mu}) \nonumber \\
 & = & ( \mathbf{y}-\mathbf{A}\bm{\mu})^T \{
 \mathbf{W}^{-1}\sigma^2 + \mathbf{A}\mathbf{W}\!_{ \beta}^{-1}\mathbf{A}^T
 \sigma_{\beta}^2\}^{-1}( \mathbf{y}-\mathbf{A}\bm{\mu}) \nonumber \\
 & = & ( \mathbf{y}-\mathbf{A}\bm{\mu})^T \bm{\Sigma}_{py}^{-1}
 ( \mathbf{y}-\mathbf{A}\bm{\mu}). \end{eqnarray} Actually, the
 term $(\mathbf{W}^{-1}\sigma^2 + \mathbf{A}\mathbf{W}\!_{ \beta}^{-1}\mathbf{A}^T
 \sigma_{\beta}^2)$ in the last line of (\ref{QuadYFinal}) is exactly equal to the
variance-covariance matrix of the random vector $(\mathbf{A}\bm{\beta}+\bm{\epsilon})$.
It is also obvious that the mean of $(\mathbf{A}\bm{\beta}+\bm{\epsilon})$ is
$\mathbf{A}\bm{\mu}$.

Inserting $F(\mathbf{y},\bm{\beta}/\sigma^2,\sigma_{\beta}^2)$ of (\ref{QuadYBeta}) and
$F_y(\mathbf{y}/\sigma^2,\sigma_{\beta}^2)$ of (\ref{QuadYFinal}) into
$m(\mathbf{y}/\sigma^2,\sigma_{\beta}^2)$ of (\ref{marginalY}) and after the integration,
we finally obtain the  marginal distribution $m(\mathbf{y}/\sigma^2,\sigma_{\beta}^2)$ of
 $\mathbf{y}$ as follows:
\beq m(\mathbf{y}/\sigma^2,\sigma_{\beta}^2) =
\frac{1}{(2\pi)^{n/2}\sqrt{\textrm{det}(\bm{\Sigma}_{py})}} \exp\left\{ -\frac{1}{2}(
\mathbf{y}-\mathbf{A}\bm{\mu})^T\bm{\Sigma}_{py}^{-1}(
\mathbf{y}-\mathbf{A}\bm{\mu})\right\}, \eeq which is the same as (\ref{marginalYFinal}).

\end{document}